\documentclass[nofootinbib,showpacs,prd]{revtex4}
\usepackage{amsmath}
\usepackage{amsfonts}
\usepackage{amssymb}
\usepackage{graphicx}
\usepackage{color}

\def\be{\begin{equation}}
\def\ee{\end{equation}}
\def\bea{\begin{eqnarray}}
\def\eea{\end{eqnarray}}

\begin{document}

\title{Cosmographic study of the universe's specific heat: A landscape for Cosmology?}

\author{Orlando Luongo$^{1,2,3}$ and Hernando Quevedo$^{1,4,5}$}\email{orlando.luongo@na.infn.it, quevedo@nucleares.unam.mx}

\affiliation{
$^1$Instituto de Ciencias Nucleares, Universidad Nacional Aut\'onoma de M\'exico, AP 70543, M\'exico, DF 04510, Mexico.\\
$^2$Dipartimento di Scienze Fisiche, Universit\`a di Napoli "Federico II",
             Via Cinthia, I-80126, Napoli, Italy.\\
$^3$INFN Sez. di Napoli, Compl. Univ. Monte S. Angelo Ed. N
Via Cinthia, I- 80126, Napoli, Italy.\\
$^4$Dipartimento di Fisica and Icra, Universit\`a di Roma "La Sapienza", I-00185 Roma, Italy.\\
$^5$Instituto de Cosmologia, Relatividade e Astrofisica ICRA - CBPF\\
Rua Dr. Xavier Sigaud, 150, CEP 22290-180, Rio de Janeiro, Brazil.
}

\begin{abstract}
We propose a method for constructing the specific heat for the
universe by following standard definitions of classical thermodynamics,
in a spatially flat homogeneous and isotropic spacetime. We use cosmography
to represent the specific heat in terms of measurable quantities, and show
that a negative specific heat at constant volume and a zero specific heat at
constant pressure are compatible with observational data.  We derive the
most general cosmological model which is compatible with the values obtained
for the specific heat of the universe, and show that it overcomes the
fine-tuning and the coincidence problems of the $\Lambda$CDM model.
\end{abstract}


\maketitle

\section{Introduction}
\label{sec:int}

All the attempts made to understand the modern cosmological observations
indicate the existence of a positive acceleration that has been
extensively measured in different surveys \cite{sn0,sn1,sn2,sn3,sn4,sn5}.
On the other hand, General
Relativity predicts a decelerated expansion, if one assumes that the
universe is filled by dust only \cite{JorgeSmoot}. Thus,
the universe turns out to be dominated by a new
ingredient that is responsible for the cosmic acceleration \cite{rev}.
Unfortunately, this ingredient forecasts a both surprising and
unexpected negative Equation of State (EoS) \cite{v1,v2,v3}.
The physical interpretation of this exotic behavior
reflects the need of counterbalancing the gravitational attraction
\cite{proc}. Due to this anomalous behavior, the unknown
term to be added in the cosmological puzzle is usually referred to
as Dark Energy (DE), filling almost the 70$\%$ of the whole
universe \cite{bston}. Thence, soon after the discovery of the
anomalous dynamics of DE, the former cosmological model was forced
to be modified. Thus, thought as the likely most viable explanation,
it has been assumed that a cosmological constant $\Lambda$
\cite{lambda1,lambda2,lambda3,lambda4} should be included within the
Einstein equations. The corresponding model, the so-called
$\Lambda$CDM paradigm, depicts a cold Dark Matter and a
late-dominant counterpart represented by the density of $\Lambda$
\cite{lambdax}, i.e. $\Omega_\Lambda$. Since the model excellently
passes all the experimental tests, it strongly entered the modern
cosmological convictions, becoming the standard cosmological
paradigm \cite{tsu}.

Unfortunately, two profound issues plague the model:
the problems of coincidence and of fine-tuning \cite{coppa}.
Particularly, the coincidence points out that an impressive amount
of available cosmic data predicts a matter density, $\Omega_m$ extremely
close to $\Omega_\Lambda$. On the other hand, the fine-tuning issue
poses several limits on the forecasted cosmological constant and on
the measured one. Due to these shortcomings, many other alternative
explanations to the problem of DE have been proposed  in
literature throughout the time \cite{www,u1,u2,u3,u4}; however, a
self-consistent explanation, both from the theoretical and experimental points of view, seems to be
so far out of reach \cite{Paddyreview}.

On the other hand, many
efforts have been made recently in order to analyze the
thermodynamic properties of cosmological models by using only the geometry
of the space of equilibrium states \cite{quev07,quew}
\footnote{A different approach takes into account
quantum effects due to fluctuations with particle creation
\cite{gnegne,P61,P88,P89,D97}.}.
In particular, it turns out that negative specific heats and non-extensive
thermodynamic variables can be used to describe systems with negative EoS's \cite{abcq12}.
This kind of systems can be studied further in the context of geometrothermodynamics
\cite{quev07}, a formalism that allows one to define and handle non-extensive variables,
or by using non-extensive generalizations of the concept of entropy \cite{tallo}.

In this work, we use the standard definitions of classical thermodynamics to
define the specific heat at constant volume and at constant pressure on
a cosmological background, consisting of the Friedman-Lema\^itre-Robertson-Walker
spacetime and a perfect fluid as the source of the gravitational field.
The formalism of cosmography \cite{wei,vix,vix2,vixbis} is used to represent
the specific heat in terms of cosmographic parameters whose values are determined
indirectly from the observational data for the redshift and the luminosity
distance\footnote{The luminosity distance is not the only observable that can be used to determine
the cosmographic parameters. For a different approach see, for instance, \cite{yeah}.}
\cite{star1,star2}.
In addition, we will relate the thermodynamic properties of the universe \cite{line} with
a cosmological model which is able to explain the DE effects, using our thermodynamic assumptions only.

This paper is  organized as follows: In Sec. \ref{sec:heats}, we formulate the
definition of the specific heat at constant volume and at constant pressure and show
that the corresponding thermodynamic potentials are positive  as a consequence
of the weak energy conditions.  In Sec. \ref{sec:cosm}, we present the main
idea of cosmography and evaluate the cosmographic parameters in the case of
our universe. Sec. \ref{sec:csh} is devoted to the study of the specific heat
in terms of the cosmographic parameters.
In Sec. \ref{sec:mod}, we derive the most general cosmological model that
is compatible with the values of the specific heat of the universe.
Finally, Sec. \ref{sec:con} contains the
discussions of our results and suggestions for further research.

\section{Specific heat of the universe}
\label{sec:heats}

To describe the geometric and physical properties of the universe at large scales we
assume the validity of the cosmological principle. Then, the geometric properties
are determined by the Friedman-Lema\^itre-Robertson-Walker (FLRW) line element
\begin{equation}\label{frwmetric}
ds^2=dt^2-a(t)^2\Bigl[dr^2+r^2\left(\sin^2\theta
d\phi^2+d\theta^2\right)\Bigl]\, ,
\end{equation}
where, in accordance with observations, we assume a zero spatial curvature.
For a perfect fluid source, the Einstein equations reduce to the Friedman equations
\be
\label{nrk}
H^2 = \frac{8\pi G}{3}\rho_t\,, \quad
\dot H + H^2=-\frac{4\pi G}{3} \Bigl(3P+\rho_t\Bigr)\,,
\ee
where the dots represent the derivative with respect to the cosmic time $t$.

The perfect fluid is the source of the gravitational field of the universe at
large scales, and can also be interpreted intuitively as a thermodynamic system \cite{mtw}.
Moreover, it can be shown that the  laws of thermodynamics are mathematically consistent
with an isotropic and homogeneous geometric background \cite{herny}. In fact, using
the first law of thermodynamics one can derive an expression for the temperature of
the universe whose predictions are in agreement with the observations of the cosmic
background radiation \cite{boul,boul2,paddy,bjo}. Then, following standard definitions of classical thermodynamics \cite{kundu},
we introduce the specific heat
\begin{equation}\label{sph}
\textsf{C}\equiv\frac{\partial \hat U}{\partial T}\,,
\end{equation}
where the energy function $\hat U$ determines the type of specific heat.
Indeed, according to the
standard definitions  of internal energy $U$ and
enthalpy $h$, the following relations hold
\begin{eqnarray}
\textsf{C}_V = \frac{\partial U}{\partial T}\,,\quad\quad
\textsf{C}_P = \frac{\partial h}{\partial
T}\,,
\end{eqnarray}
representing the specific heat at constant pressure $P$ and constant volume $V$, respectively.

For the FLRW model, the energy and enthalpy of the universe can be computed as
\be
\label{u}
U= V_0\rho_t a^{3}\,,\quad h =  V_0\left(\rho_t+P\right)a^3\, ,
\ee
where we considered the evolution of the volume up to the first order
as $V=V_0a^3$, with $V_0$ being at least proportional to the universe's radius $\frac{1}{H_0}$.
Furthermore, if we assume the weak energy conditions $\rho_t\geq 0$ and $\rho_t+P \geq 0$, it follows that
$U$ and $h$ are positive quantities\footnote{Notice that, despite the total density must be
positive, the individual species may be negative, without violating
the weak energy conditions.}. In general, the assumption $V=V_0a^{3}$ fulfills the weak energy condition, although other approaches suggest that
the volume can be function of the apparent horizon, i.e. $r\propto H^{-1}$ \cite{aggiungiamo1}. The choice $V\propto r^{3}=V_0\,H^{-3}$ would reproduce a causal region with the entropy proportional to $H^{-2}$. However, to characterize the specific heats, we are interested in the small redshift regime, i.e. $z\ll 1$. Thus, under the limits of small $z$, one can first approximate the volume with the simplest choice of $V\propto a^{3}$, without contradicting the causality condition \cite{aggiungiamo2}.

Our goal is to analyze the time dependence of the specific heat. For comparisons with observational data, however,
it is more convenient to use the redshift $z$ which is defined by means of the standard relationships $a = \frac{1}{1+z}$ and
$\frac{dz}{dt} = - (1+z)H$. Then, $\textsf{C} = \frac{d \tilde U}{dz}\left(\frac{\partial T}{dz}\right)^{-1}$, and using Eq.(\ref{u}) and
the Friedman equations,  we obtain
\begin{equation}\label{uzaaa}
\textsf{C}_V = \frac{1}{T'}
\frac{dU}{dz}=\frac{3V_0a^3}{8\pi
GT'}\left(\frac{dH^2}{dz}-3a\,H^2\right)\,,
\end{equation}
and
\begin{equation}\label{hsxxx}
\textsf{C}_P=\frac{1}{T'}
\frac{dh}{dz}=\frac{1}{T'}\frac{dU}{dz}+\frac{V_0a^3}{T'}\left(\frac{dP}{dz}-3a\,P\right)\,,
\end{equation}
where the prime denotes differentiation with respect to the redshift parameter $z$.

\section{Cosmography of the universe}
\label{sec:cosm}

The aim of cosmography is the study of the kinematic
quantities that characterize the cosmological scenario, without
using any particular theory that dictates the dynamics of the
gravitational field.  For this reason, cosmography is also called cosmokinetics,
or kinematics of the universe. To develop the formalism of cosmography, it is necessary to specify only a particular
model for the spacetime. If we assume the FLRW model, it is clear that the only the scale factor must be considered.
Then, the Taylor expansion of $a(t)$ can be written as follows \cite{jumz}
\be
\label{serie1}
a(t)  =  a_0 \Bigl[ 1+ H_0 \Delta t - \frac{q_0}{2} H_0^2 \Delta
t^2 + \frac{j_0}{6} H_0^3 \Delta t^3 +\ldots\bigr]\,,
\ee
where the subscript $0$ indicates that the corresponding parameters are
are evaluated at $z=0$. The deceleration $q$ and jerk $j$ are defined
as
\be
\label{pinza}
q = -\frac{d^2a}{dt^2}\left(H^2a\right)^{-1}\,, \quad j
=\frac{d^3a}{dt^3}\left(H^3a\right)^{-1}\,.
\ee
Alternatively, one can also use the Hubble parameter to obtain
\be
\label{parametri}
\frac{\dot{H}}{H^2}=-\left(q + 1\right)\,, \qquad \frac{\ddot{H}}{H^3} = 2+j+3q\,.
\ee
Using Eqs. ($\ref{serie1}$) and ($\ref{pinza}$), it is possible to
expand the luminosity distance
\begin{equation}\label{dl1000}
d_l=(1+z)\int_{0}^{z}\frac{d\upsilon}{H(\upsilon)}\,,
\end{equation}
in terms of the redshift $z$, as follows
\begin{eqnarray}
\label{zump}
d_L & = & \frac{1}{H_0} \Bigl[ z + \frac{1}{2}\Bigl(1-q_0
\Bigr)z^2+\frac{1}{6}\Bigl(3q_0^2+q_0-j_0-1\Bigr)z^3+\ldots\Bigr]\,.
\end{eqnarray}

Since the observational data includes the redshift and the luminosity distance,
one can now use the cosmographic series (\ref{zump}) to determine the values
of the remaining parameters. To this end, we perform a chi-squared analysis,
using the  most recent Union 2.1 compilation \cite{kow}. We first carry out an
analysis in which the parameters to be determined, i.e., $H_0,q_0$ and $j_0$,
are not confined in any particular interval. In the second analysis, we fix
the Hubble constant by using the value obtained from WMAP 7 \cite{kow}, i.e.,
$H_0=70.2 Km\,s^{-1}\,Mpc^{-1}$. The results of the first and the second
analysis are presented in Tabs. I and II, respectively.

\begin{table}
\begin{center}\label{tab:I}
\begin{tabular}{|c|c|c|c|}
\hline
$H_0\equiv H(z=0)$  &  $q_0\equiv q(z=0)$ & $j_0\equiv j(z=0)$ \\
\hline
\hline
&&\\
$70.50^{+1.60}_{-1.60}$ & $-0.628_{-0.082}^{+0.085}$ & $0.979^{+0.531}_{-0.563}$ \\
&&\\
\hline
\hline
\end{tabular}
\caption{Summary of the numerical results obtained by using Eq.
($\ref{zump}$), with  the most recent Supernovae Ia data
contained in the Union 2.1 compilation. $H_0$ is given in units $s^{-1}Km\,Mpc^{-1}$.
For these values we obtain $\chi^2=0.975$.}
\end{center}
\end{table}

\begin{table}
\begin{center}\label{tab:II}
\begin{tabular}{|c|c|c|c|}
\hline
$H_0\equiv H(z=0)$  &  $q_0\equiv q(z=0)$ & $j_0\equiv j(z=0)$ \\
\hline
\hline
&&\\
$70.30^{+1.40}_{-1.40}$ & $-0.595^{+0.089}_{-0.085}$ & $0.816^{+0.539}_{-0.572}$ \\
&&\\
\hline
\hline
\end{tabular}
\caption{Summary of the numerical results obtained by using Eq.
($\ref{zump}$), with the fixed value $H_0=(70.30\pm1.40)\,s^{-1}Km\,Mpc^{-1}$, as obtained from WMAP 7 \cite{kow}. In this case, $\chi^2=0.973$.}
\end{center}
\end{table}

\begin{figure}[ht]
\includegraphics*[scale=0.5]{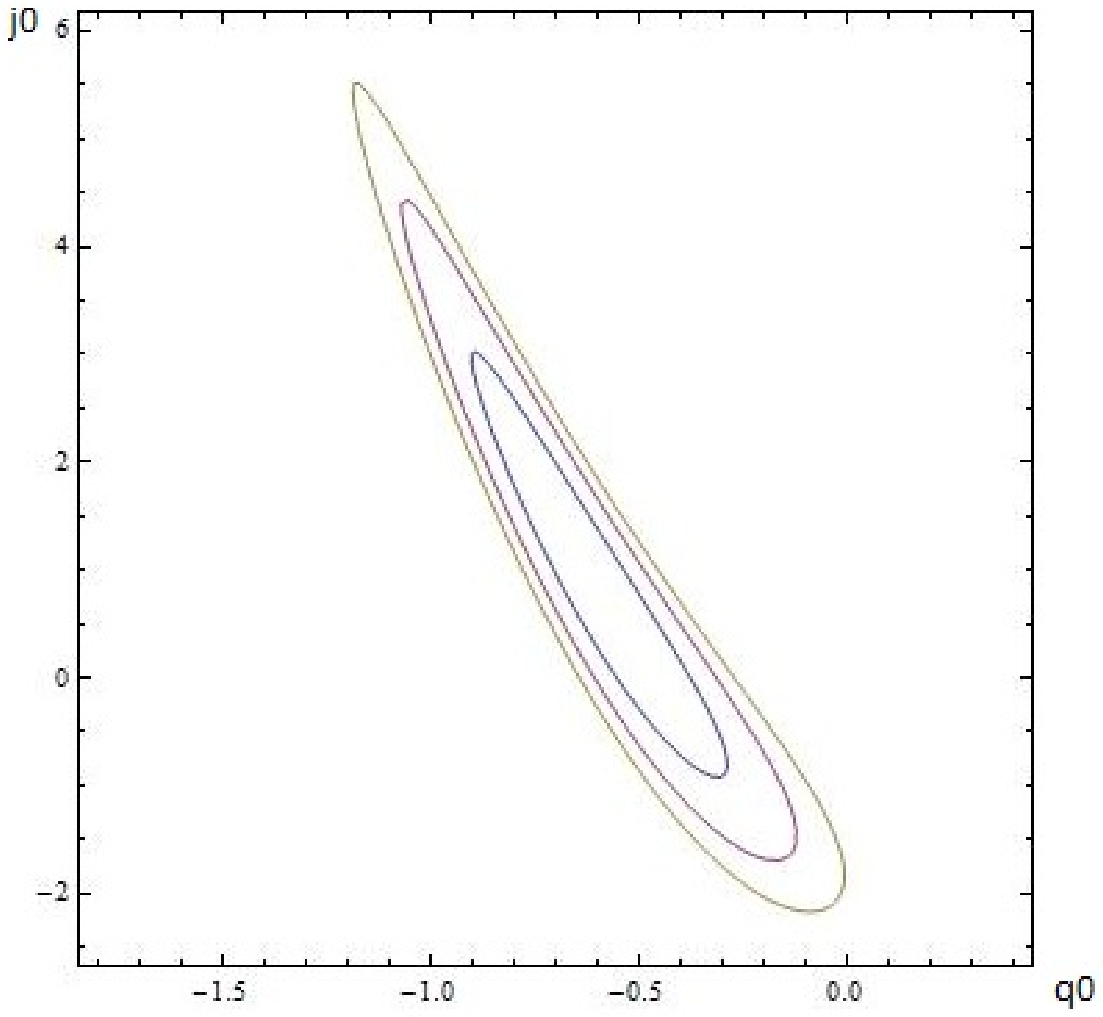}
\caption{Contour plots of $j_0$ VS $q_0$ at one, two and three sigma of confidence levels for the fitting results of Tab. I.}\label{3q1}
\end{figure}

\begin{figure}[ht]
\includegraphics*[scale=0.5]{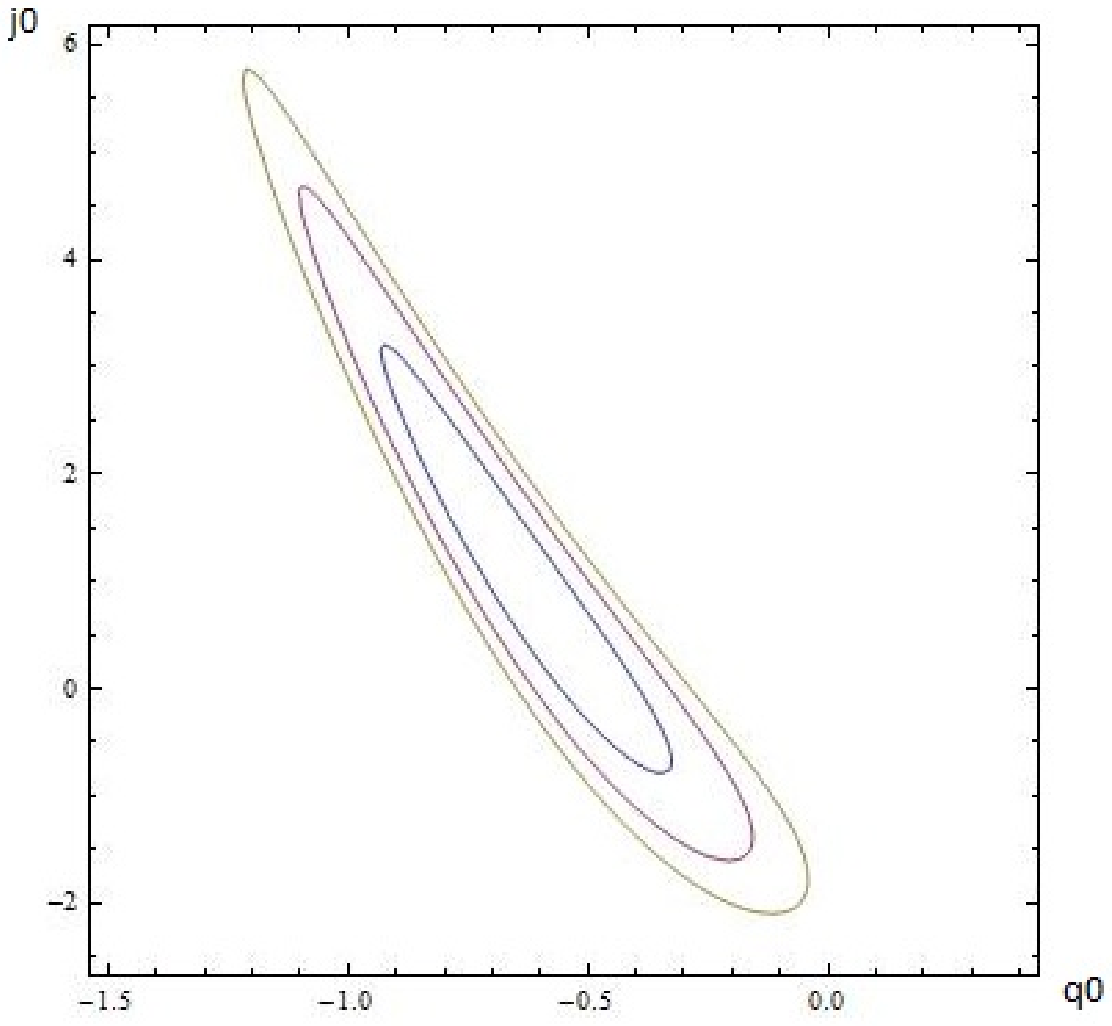}
\caption{Contour plots of $j_0$ VS $q_0$ at one, two and three sigma of confidence levels for the fitting results of Tab. II.}\label{3q2}
\end{figure}

Notice that the results we obtain for $q_0$ are compatible with the results presented in previous works \cite{v1,v2}.
Moreover, the result for $j_0$, suggesting that  $j_0\sim 1$ \cite{v3,luongo}, is in accordance
with the exact theoretical bound predicted by the $\Lambda$CDM model. In the next section, we will show that the values
of the cosmographic parameters can be used to limit the values of the specific heat.

\section{Cosmographic analysis of the specific heat}
\label{sec:csh}

To find the cosmographic series of the heat capacities, we first
rewrite Eqs.
($\ref{parametri}$) as
\begin{equation}\label{hz1}
\frac{dH}{dz}=H\frac{1+q}{1+z}\,,\quad\frac{d^2H}{dz^2}=H\frac{j-q^2}{(1+z)^2}\,,
\end{equation}
where we used $a=\frac{1}{1+z}$ and $\frac{dz}{dt}=-H(1+z)$.
Moreover, in terms of $z$, the continuity  the continuity equation in a FLRW universe can be expressed as
\begin{equation}\label{conty}
\frac{1}{3}\frac{d\rho_t}{dz}=\frac{P+\rho_t}{1+z}\,.
\end{equation}

Using Eqs. ($\ref{conty}$) and  ($\ref{nrk}$) in the expressions for the specific heats (\ref{uzaaa}) and (\ref{hsxxx}), we finally
obtain
\begin{eqnarray}\label{bouh}
\textsf{C}_P&=&\frac{V_0}{4\pi G}\frac{H^2}{T^{'}}\frac{j-1}{(1+z)^4}\,,\nonumber\\
\,\\
\textsf{C}_V&=&\frac{V_0}{8\pi G}\frac{H^2}{T^{'}}\frac{2q-1}{(1+z)^4}\,,\nonumber
\end{eqnarray}
where we made use of $\frac{dP}{dz}=\frac{H^2}{4\pi G}\frac{j-1}{1+z}$.
Eqs. ($\ref{bouh}$) show the direct measurement of the cosmographic parameters
$H$, $q$ and $j$, at a given redshift, implies the indirect measurement of
the specific heats. In particular, for a measurement at $z=0$, we have that
\begin{eqnarray}\label{iosis}
  \textsf{C}_{P0} &=& \frac{V_0}{4\pi G}\frac{H_{0}^{2}}{T^{'}_{0}}(j_0-1)\,,\nonumber\\
  \, \\
  \textsf{C}_{V0} &=& \frac{V_0}{8\pi G}\frac{H_{0}^{2}}{T^{'}_{0}}(2q_0-1)\,.\nonumber
\end{eqnarray}
First, notice that the sign of $\textsf{C}_{P0}$ is determined by the sign of the expression $j_0-1$. Moreover,
we can see that a positive value of $\textsf{C}_{V0}$ implies a decelerated universe. Therefore, to obtain a universe that is currently in an
accelerated phase, we are forced to assume that $\textsf{C}_V<0$.

Using the method of standard error propagation, from Eqs. (\ref{iosis}) we infer the numerical bounds reported in Tab. III.
\begin{table}[h]
\begin{center}\label{tab:III}
\begin{tabular}{|c|c|c|c|}
\hline
$\textsf{C}_{P0}\equiv\textsf{C}_P(z=0)$  &  $\textsf{C}_{V0}\equiv\textsf{C}_V(z=0)$  \\
\hline
\hline
&\\
$-0.030^{+0.748}_{-0.793}$ & $-1.587^{+0.156}_{-0.151}$ \\
&\\
\hline
\hline
\end{tabular}
\caption{Summary of the numerical results obtained from Eq.
($\ref{iosis}$). We used $G=6.67\times10^{-11}Nm^2Kg^{-2}$, $V_{0}\equiv\left(\frac{1}{H_0}\right)^{3}$ and conventionally
$T^{'}\equiv 1K$. The error propagation is logarithmic. For $q_0$ and $j_0$ and $H_0$, we used the values of Tab. I, and we inserted back the explicit numerical value of the light velocity so that $\textsf{C}_{P0}$ and $\textsf{C}_{V0}$ are expressed in units $\frac{J}{K}$. The specific heats are expressed in powers of $10^{52}$.}
\end{center}
\end{table}
We see that the results are
compatible  with a zero value of $\textsf{C}_{P0}$ and with a negative $\textsf{C}_{V0}$. These values are also supported by
the condition $\textsf{C}_P-\textsf{C}_V>0$, which is essential to guarantee the stability of a  thermodynamical system.
Notice that to obtain the numerical values presented in Tab. III we assume that $T'=1$. It is clear that $T'$ must be positive to be in agreement with
the temperature evolution since the epoch after the big bang. The explicit value of $T'>0$ does not affect the main results of our analysis
which are a negative value for $\textsf{C}_V$ and a value close to zero for $\textsf{C}_P$.

It is important to mention that the above numerical bounds can be confirmed in the $\Lambda$CDM model.
Indeed, at $z=0$ the $\Lambda$CDM model
predicts \cite{yeah}
\be
\label{lcdmcosm}
q_{0\,\Lambda CDM} = -1+\frac{3}{2}\Omega_m\ ,\quad j_{0\,\Lambda CDM} = 1 .
\ee
Moreover, the jerk turns out to be fixed as $j=1$ for all values of $z$. Consequently,
in the range $\Omega_m=0.274\pm0.015$, $\Lambda$CDM predicts that
$\textsf{C}_P=0$ and $\textsf{C}_V<0$. Thus, we can conclude that our cosmographic analysis
of the specific heat of the universe is in agreement with the predictions of the $\Lambda$CDM model.
Furthermore, an interesting question arises: Is $\Lambda$CDM the most general model that is compatible
with the specific heats $\textsf{C}_P=0$ and $\textsf{C}_V<0$?

We conclude that the behavior  of the specific heat of the universe may suggest insights for understanding
the cosmological landscape. In the next
section, we will study the consequences of a vanishing
$\textsf{C}_P$ for all $z$ in the context of relativistic cosmology.

\section{Cosmology with vanishing specific heat}
\label{sec:mod}

According to the results of Secs. III and IV, the present value of the jerk parameter is compatible with $j_0=1$. This is in accordance with the cosmological observations and seems to confirm the $\Lambda$CDM paradigm, which predicts $j=1$ for all the values of the redshift $z$. However, $\Lambda$CDM is not the only model that can reproduce this result. Indeed, as a direct consequence of the first of Eqs. ($\ref{bouh}$), the most general perfect fluid, which reproduces the constraint $j=1$, satisfies the condition $w\propto \rho^{-1}$. Then, by using Eq. ($\ref{conty}$), one gets
\begin{eqnarray}\label{kjnmb}
\rho(z)&=&\rho_0\big[1-\kappa(1+z)^3\big]\,,\nonumber\\
\,\\
w(z)&=&-\frac{1}{1-\kappa(1+z)^3}\,,\nonumber
\end{eqnarray}
where $\kappa$ is an integrating constant. Notice that the first equation of (\ref{kjnmb}) can be interpreted
as a thermodynamic correction to the collisionless dust term.

Under this hypothesis, the Hubble rate of the corresponding cosmological model is therefore
\begin{equation}\label{Hz2}
H=H_0\sqrt{\frac{\kappa}{\kappa-1}(1+z)^3+\frac{1}{1-\kappa}}\,,
\end{equation}
\noindent which generalizes the $\Lambda$CDM model \cite{luongo}. In fact, Eq. ($\ref{Hz2}$) reduces to $\Lambda$CDM in the limiting case $\kappa\rightarrow1-\frac{1}{\Omega_\Lambda}$, where $\Omega_\Lambda$ is the cosmological constant density. In addition, we notice that $P<0$, by requiring $\dot H>0$.

It is important to mention that although the $\Lambda$CDM model is recovered as a special case,
the interpretation of our DE term, i.e. $\Omega_X\equiv\frac{1}{1-\kappa}$, is not that of the vacuum energy. Its origin is a consequence of the vanishing of the specific heat $\textsf{C}_{P}$.

In order to fix the theoretical limits on $\kappa$ and to predict its sign, we can find the corresponding temperature evolution. So, since the universe enthalpy is a constant and the universe evolves adiabatically, for a perfect gas we have that $P=-\rho \left(\textsf{C}_V-\textsf{C}_P\right) T$,
and $
P=\tilde P\rho^\frac{\textsf{c}_P}{\textsf{c}_V}$, where $\tilde P$ is an integration constant. The corresponding EoS is therefore $w=-C_V T$ with an explicit temperature evolution forced to be
\begin{equation}\label{temperature}
T=T_0(1+z)^{3\gamma\kappa}e^{-\gamma\Bigl[1-\frac{1}{(1+z)^3}\Bigr]}\,,
\end{equation}
with $\gamma\equiv\frac{H_0^3V_0}{8\pi G}\Omega_X$ and $T_0$ the temperature at our time, i.e. $T_0=2.73 K$.

The temperature evolution, found under the hypothesis of $\textsf{C}_P=0$ and $\textsf{C}_V<0$, agrees with the thermodynamic results obtained in \cite{line}. In addition, we must assume $\kappa<0$, in order to reproduce the temperature evolution of non-relativistic particles in standard cosmology. Thus, according to Eqs. ($\ref{kjnmb}$), the sign of $\kappa$ is able to {naturally} predict an accelerated universe, without imposing any further condition on the Einstein equations. This requirement seems to confirm the need of extending the $\Lambda$CDM model as a limiting case of a more general paradigm and candidates our approach to be a serious alternative to the $\Lambda$CDM paradigm.

\section{Conclusions}
\label{sec:con}

In this work, we use standard definitions of classical thermodynamics to
define the specific heats of the universe on a cosmological background
given by the FLRW spacetime and a perfect fluid as the source of the gravitational
interaction. We use the weak energy conditions to prove the positivity of the
internal energy and the enthalpy, the thermodynamic potentials that define
the specific heat at constant volume and pressure, respectively.

We propose to use cosmography to relate the specific heats with quantities
that can be evaluated from observational data. In particular, we established
the relationship between the specific heat and the cosmographic parameters corresponding
to acceleration $q_0$ and jerk $j_0$.  Since the explicit value of the cosmographic parameters
can be determined from the observed values for the redshift and the luminosity distance,
we show that today the specific heat at constant volume $\textsf{C}_V$ is negative and the
specific heat at constant pressure $\textsf{C}_P$ is close to zero. These results are
in agreement with the $\Lambda$CDM model in which $\textsf{C}_P$ is zero and does not evolve in time.

We also analyzed the most general cosmological model which can be derived from the assumption $\textsf{C}_P=0$.
It turns out that the $\Lambda$CDM model is contained in this generalized model as a special case, and that
the known problems of the  $\Lambda$CDM paradigm can be overcome in a canonical way. In fact, although we
consider Einstein's equations without cosmological constant, the model presented here leads to an emerging
cosmological constant, which, therefore,  has no reason to be associated
to the  vacuum energy. This fact naturally avoids the fine-tuning problem,
because in our model there is no cosmological constant in the Einstein equations that should
be compared with the predictions of Quantum Field Theory.
Concerning the coincidence problem, since we assume an evolutionary barotropic
factor $w = w(z)$, the corresponding value at $z=0$ is fixed my the matter evolution itself.
The fixing of this initial condition naturally avoids the coincidence problem; in fact,
it is not casual that we live in a time in which the energy density, which explains the acceleration of
the universe, has a magnitude comparable with $\Omega_m$.

Thus, we conclude that it represents a viable
alternative to the $\Lambda$CDM approach. However, it is essential to perform additional  tests to show that
it can be used as a realistic cosmological alternative. In particular, in the future we expect to perform
a detailed analysis of the consequences of our approach on structure formation and on the evolution of the universe through different epochs.

\section*{Acknowledgements}

We are grateful to S. Capozziello, D. Pav\'on and A. Bravetti for
fruitful discussions. This work was supported in part by DGAPA-UNAM, grant No. 106110, and Conacyt, grant No. 166391.


\begin{thebibliography}{}

\bibitem{sn0}
S. Perlmutter, et al., ApJ, {\bf 483}, 565, (1997); S. Perlmutter,
et al., Nature, {\bf 391}, 51, (1998).

\bibitem{sn1}
A. G. Riess et al., Astr. Phys. Jour., {\bf 116}, 1009, (1998).

\bibitem{sn2}
S. Perlmutter et al., Astr. Phys. Jour., {\bf 517}, 565, (1999).

\bibitem{sn3}
R. Rebolo et al., Mon. Not. Roy. Astr. Soc., {\bf 353}, 747, (2004)

\bibitem{sn4}
A. C. Pope et al., Astr. Phys. Jour., {\bf 607}, 655, (2004)

\bibitem{sn5}
P. McDonald et al., eprint: ArXiv:astro-ph/0405013, (2004).

\bibitem{JorgeSmoot}
J. L. Cervantes-Cota, G. Smoot, AIP Conf. Proc. Vol., {\bf 1396}, 28-52, (2011).

\bibitem{rev}
M. Li, X.D. Li, S. Wang, Y. Wang, Com. Theor. Phys., {\bf 56}, 525-604, (2011).

\bibitem{v1}
T. Padmanabhan, Phys. Rept., {\bf 380}, 235, (2003).

\bibitem{v2}
M. Tegmark et al. (SDSS), Phys. Rev. D, {\bf 74}, 123507, (2006).

\bibitem{v3}
W. J. Percival et al., Mon. Not. Roy. Astr., Soc., {\bf 381}, 1053, (2007).

\bibitem{proc}
O. Luongo, H. Quevedo, ArXiv: gr-qc/1005.4532, (2010).

\bibitem{bston}
S. Weinberg, Cosmology, Oxford Univ. Press, Oxford, (2008).

\bibitem{lambda1}
S.M. Carroll, W.H. Press, E.L. Turner, ARAA, {\bf 30}, 499, (1992).

\bibitem{lambda2}
V. Sahni, A. Starobinski, Int. J. Mod. Phys. D, {\bf 9}, 373, (2000).

\bibitem{lambda3}
M. Tegmark et al., Phys. Rev. D, {\bf 69}, 103501, (2003).

\bibitem{lambda4}
U. Seljak et al., Phys. Rev. D, {\bf 71}, 103515, (2005).

\bibitem{lambdax}
S.~M.~Carroll, Liv. Rev. Rel., {\bf 4}, 1, (2001).

\bibitem{tsu}
S. Tsujikawa, eprint: ArXiv:gr-qc/1004.1493, (2010).

\bibitem{coppa}
J. E. Copeland, M. Sami, S. Tsujikawa, Int. J. Mod. Phys. D, {\bf
15}, 1753-1936, (2006).

\bibitem{www}
S. Weinberg, Rev. Mod. Phys., {\bf 61}, 1, (1989).

\bibitem{u1}
R. Durrer, R. Maartens, Gen. Rel. Grav., {\bf 40}, 301-328, (2008);
V. Sahni, A. A.~Starobinsky, Int. J. Mod. Phys. D, {\bf 9}, 373,
(2000); V. Sahni, Lect. Not. Phys., {\bf 653}, 141, (2004).

\bibitem{u2}
P. J. E. Peebles, B. Ratra, Rev. Mod. Phys.,  {\bf 75}, 559, (2003);
C. Wetterich, Nucl. Phys. B., {\bf 302}, 668, (1988); Y. Fujii, T.
Nishioka, Phys. Rev. D, {\bf 42}, 361, (1990).

\bibitem{u3}
S. M. Carroll, Phys. Rev. Lett.,  {\bf 81}, 3067, (1998); E. J.
Copeland, A. R. Liddle, D. Wands, Phys. Rev. D, {\bf 57}, 4686,
(1998).

\bibitem{u4}
R. R. Caldwell, R. Dave and P. J. Steinhardt, Phys. Rev. Lett., {\bf
80}, 1582, (1998); I. Zlatev, L. M. Wang, P. J. Steinhardt, Phys.
Rev. Lett., {\bf 82}, 896, (1999); P. J. Steinhardt, L. M. Wang, I.
Zlatev, Phys.  Rev.  D, {\bf 59}, 123504, (1999).

\bibitem{Paddyreview}
T. Padmanabhan, Phys. Rept., {\bf 380}, 235, (2003).

\bibitem{quev07} H. Quevedo, J. Math. Phys. {\bf 48} 013506 (2007).

\bibitem{quew}
A. Vazquez, H. Quevedo, A. Sanchez, J. Geom. Phys., {\bf 60}, 1942,
(2010).


\bibitem{gnegne}
G. Montani, Class. Quant. Grav., {\bf 18}, 193-203, (2001).


\bibitem{P61}
I. Prigogine, \emph{Thermodynamics of Irreversible Processes},
(Wiley New York), (1961).

\bibitem{P88}
I. Prigogine et al., Proc. Nat. Acad., (1988).

\bibitem{P89}
I. Prigogine et al., Gen. Rel. and Grav., {\bf 21}, 767, (1989).

\bibitem{D97}
K. Desikan, Gen. Rel. and Grav.,  {\bf 29}, 4, 435, (1997).

\bibitem{abcq12}
A. Aviles, A. Bastarrachea, L. Campuzano and H. Quevedo,
Phys. Rev. D, {\bf 86}, 063508, (2012).

\bibitem{tallo}
C. Tsallis, \emph{Non-extensive Statistical Mechanics and
Applications}, (eds. S. Abe and Y. Okamoto), Springer, 3-98, (2001).

\bibitem{wei}
S. Weinberg, \emph{Gravitation and Cosmology: Principles and
applications of the general theory of relativity}, (Wiley, New
York), (1972).

\bibitem{vix}
M. Visser, C. Catto$\ddot{e}$n, Class. Quant. Grav., {\bf 24}, 5985,
(2007)

\bibitem{vix2}
M. Visser, C. Catto$\ddot{e}$n, ArXiv:gr-qc/0703122, (2007).

\bibitem{vixbis}
U. Alam, V. Sahni, T. D. Saini, and A. A. Starobinsky, Mon. Not.
Roy. Astr., Soc., {\bf 344}, 1057, (2003).

\bibitem{yeah}
O. Luongo, Mod. Phys. Lett. A, {\bf 26}, 20, (2011); A. Aviles, L. Bonanno, O. Luongo, H. Quevedo, Phys. Rev. D, {\bf 84}, 103520, (2011); A. Aviles, C. Gruber, O. Luongo, H. Quevedo, Arxiv: 1204.2007, (2011).

\bibitem{star1}
V. Sahni, T. D. Saini and A. A. Starobinsky, JETP Lett., {\bf 77},
201, (2003).

\bibitem{star2}
U. Alam, V. Sahni, T. D. Saini, and A. A. Starobinsky, Mon. Not.
Roy. Astr., Soc., {\bf 344}, 1057, (2003).

\bibitem{line}
R. Silva, R. S. Goncalves, J. S. Alcaniz, H. B. Silva, Astr. and Astroph., {\bf 537}, A11, (2012); J. C. Carvalho, J. S. Alcaniz, Month. Not. R. Astr. Soc., {\bf 418}, 1873, (2011).

\bibitem{mtw}
J. A. S. Lima, J. S. Alcaniz, Phys. Lett. B, {\bf 600}, 191, (2004); C. W. Misner, K. S. Thorne, J. A. Wheeler, \emph{Gravitation}, W. H. Freeman, (1973).

\bibitem{herny}
A. Krasinski, H. Quevedo, R. Sussman, J. Math. Phys., {\bf 38},
2602, (1997).

\bibitem{boul}
D. Lynden-Bell and R. Wood, Mon. Not. Roy. Astr., Soc., {\bf 138},
495, (1968).

\bibitem{boul2}
D. Lynden-Bell, Physica A, {\bf 263}, 1-4, (1999).

\bibitem{paddy}
T. Padmanabhan, Phys. Rep., {\bf 188}, 285, (1990).

\bibitem{bjo}
B. Einarsson, Phys. Lett. A, {\bf 332}, 335-344, (2004).

\bibitem{kundu}
P. K. Kundu, I. M. Cohen, \emph{Fluid Mechanics}, Elsevier Acad.
Press, San Diego, USA, (2004).

\bibitem{aggiungiamo1}
R. G. Cai, S. P. Kim, JHEP, 0502, 050, (2005); R. G. Cai, L. M. Cao, Y. P. Hu, Class. Quant. Grav., 26, 155018, (2009); S. del Campo, I. Duran, R. Herrera, D. Pavon, Phys. Rev. D, 86, 083509, (2012).

\bibitem{aggiungiamo2}
T. Jacobson, Phys. Rev. Lett., 75, 1260, (1995); C. Rovelli, arXiv:1209.0065, (2012).

\bibitem{jumz}
L. Xu, Y. Wang, Phys. Lett. B, {\bf 702}, 114-120, (2011).

\bibitem{kow}
N. Suzuki, D. Rubin, C. Lidman, G. Aldering, R. Amanullah, et al., Astrophys. J. {\bf 746}, 85, (2012); E. Komatsu, et al., Astrop. J. Sup., {\bf 192}, 18, (2011).

\bibitem{luongo}
O. Luongo, H. Quevedo, ArXiv:gr-qc/1104.4758, (2011); O. Luongo, H. Quevedo, Astroph. and Sp. Sci. {\bf 338}, 2, 345-349, (2011).

\end{thebibliography}
\end{document}